\documentclass[9pt, conference]{IEEEtran}
\IEEEoverridecommandlockouts

\usepackage{cite}
\usepackage{amsmath,amssymb,amsfonts}
\usepackage{algorithmic}
\usepackage{graphicx}
\usepackage{textcomp}
\usepackage{xcolor}
\usepackage{hyperref}
\usepackage{siunitx}
\usepackage{graphicx}
\usepackage{booktabs}
\usepackage{makecell}
\usepackage{multirow} 
\usepackage{caption}
\usepackage{url}
\def\BibTeX{{\rm B\kern-.05em{\sc i\kern-.025em b}\kern-.08em
    T\kern-.1667em\lower.7ex\hbox{E}\kern-.125emX}}

\begin{document}

\title{Simulating Native Speaker Shadowing for Nonnative Speech Assessment with Latent Speech Representations}

\author{\IEEEauthorblockN{Haopeng Geng, Daisuke Saito, Nobuaki Minematsu}
\IEEEauthorblockA{\textit{Graduate School of Engineering} \\
\textit{The University of Tokyo} \\
Tokyo, Japan \\
\{kevingenghaopeng, dsk\_saito, mine\}@gavo.t.u-tokyo.ac.jp}
% \and
% \IEEEauthorblockN{Daisuke Saito}
% \IEEEauthorblockA{\textit{Graduate School of Engineering} \\
% \textit{The University of Tokyo}\\
% Tokyo, Japan  \\
% dsk\_saito@gavo.t.u-tokyo.ac.jp}
% \and
% \IEEEauthorblockN{Nobuaki Minematsu}
% \IEEEauthorblockA{\textit{Graduate School of Engineering} \\
% \textit{The University of Tokyo}\\
% Tokyo, Japan  \\
% mine@gavo.t.u-tokyo.ac.jp}
}

\maketitle

\begin{abstract}
Evaluating speech intelligibility is a critical task in computer-aided language learning systems. Traditional methods often rely on word error rates (WER) provided by automatic speech recognition (ASR) as intelligibility scores. However, this approach has significant limitations due to notable differences between human speech recognition (HSR) and ASR. A promising alternative is to involve a native (L1) speaker in shadowing what nonnative (L2) speakers say. Breakdowns or mispronunciations in the L1 speaker’s shadowing utterance can serve as indicators for assessing L2 speech intelligibility. In this study, we propose a speech generation system that simulates the L1 shadowing process using voice conversion (VC) techniques and latent speech representations. Our experimental results demonstrate that this method effectively replicates the L1 shadowing process, offering an innovative tool to evaluate L2 speech intelligibility. Notably, systems that utilize self-supervised speech representations (S3R) show a higher degree of similarity to real L1 shadowing utterances in both linguistic accuracy and naturalness\footnote{Audio samples are available at: \url{https://secondtonumb.github.io/publication_demo/ICASSP_2025/index.html}}.

\end{abstract}

\section{Introduction}
Nonnative (L2) speakers often face challenges in receiving detailed, word-by-word feedback on their speaking performance, particularly when educational resources are limited. Recently, the rise of large language models (LLMs) has made computer-assisted language learning (CALL) systems more accessible, allowing L2 speakers to have their writing proofread with detailed suggestions. However, the automatic speech recognition (ASR) systems used in LLM agents are typically designed to speculate what the speakers said, which contradicts the purpose of computer assisted pronunciation training (CAPT) systems. In the worst-case scenario, a lacking of feedback on L2 speaker's speech can reinforce incorrect pronunciation patterns, leading to fossilization.

How can we effectively identify unintelligible parts in L2 speech? One promising method is native (L1) speaker shadowing, where an L1 speaker repeats an L2 utterance with minimal delay while listening. This human-involved approach to assess L2 speech intelligibility has been explored in previous research \cite{inoue18_interspeech, lin2020shadowability, zhu2021multi}. In this approach, the L1 speaker immediately repeats what he/she has heard in the given L2 speech with their own accent, thereby indicates where listening breakdowns occur. These breakdowns highlight problematic areas in the L2 speech and helping L2 speakers understand where their speech may be difficult to understand. 

Since it is impractical to provide a real shadower for every L2 speaker. In this study, our objective is to develop a virtual shadowing system that simulates the process of an L1 speaker shadowing L2 speech (L1-shadowing-L2). This system aims to offer a novel CALL tool that provides more comprehensible feedback for L2 speakers. Our contributions are as follows:

\begin{itemize}
    \item We are the first to use unique semi-parallel L1-shadowing-L2 data to develop a CAPT system that offers intuitive speech-based feedback instead of traditional statements or scores. 
    \item We employ self-supervised speech feature to present L2 speech, significantly improving the quality of the generated shadowing-like speech. Experimental evaluations demonstrate the generated speech closely resembles real shadowing in linguistic similarity and naturalness.
    \item We propose two fine-tuning approaches that utilize different types of L1 shadowing instances. Experimental results indicate that these techniques can enhance mapping performance to practical L1 shadowing speech, compared to the standard one-step training.
    \item Our study expands the application of voice conversion technology into new areas. We demonstrate that a parallel-VC setting between L1 and L2 can be effectively utilized to create a virtual shadowing system.

\end{itemize}

\section{Research background}
\begin{figure}[t]
	\centering
	\includegraphics[width=\columnwidth]{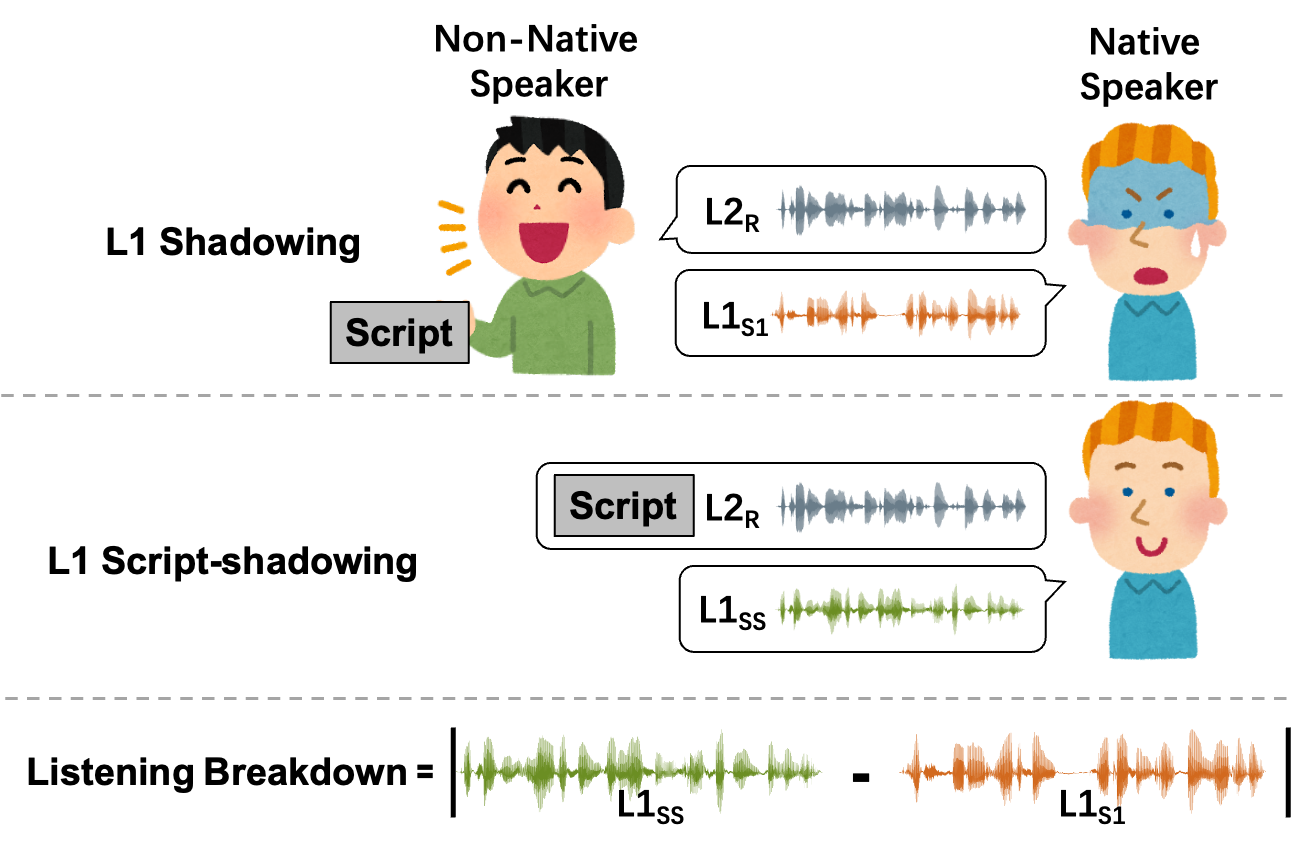}
	\captionof{figure}{Native speaker shadowing for non-native speech assessment. $L1_{S1}$ represents a native speaker’s initial shadowing, while $L1_{SS}$ denotes his/her script-shadowing, which is the most fluent shadowing.}
 %this technique aims to identify unintelligible parts in an L2 speaker's reading aloud utterances ($L2_{R}$). $L1_{S1}$ represents a native speaker’s initial shadowing, while $L1_{SS}$ denotes his/her script-shadowing, which is the most fluent shadowing. By calculating the distance between $L1_{S1}$ and $L1_{SS}$ sequencially, it is possible to pinpoint the native speaker's listening breakdowns, which correspond to the unintelligible parts in the $L2_{R}$ as well.

	\label{fig:l1shadowingl2}
\vspace{-5mm}
\end{figure} 

\subsection{Native Speech Shadowing}
Fine-grained annotation of L2 speech is such a challenging task that it requires specialist knowledge in phonetics for phoneme-level labeling. To address this problem, \cite{lin2020shadowability} proposed a two-stage reverse form \footnote{In the field of second language acquisition (SLA), shadowing typically involves learners repeating L1 speech to improve their listening skills. However, in our study, learners are shadowed by L1 raters, which we refer to as reverse shadowing.}  of shadowing to identify unintelligible parts in L2 utterances.
As shown in Fig.~\ref{fig:l1shadowingl2}, an L1 speaker first shadows a given L2 utterance alone ($L1_{S1}$). During this process, unintelligible parts result in stuttering or inarticulate production by the L1 speaker.
Following  $L1_{S1}$, the L1 speaker performs script-shadowing ($L1_{S2}$), where the L2 script is presented visually, and the L2 utterance is provided acoustically. By comparing $L1_{S1}$ and $L1_{SS}$ using dynamic time wrapping (DTW), we obtain sequential data on broken articulation—shadowing disfluencies that indicate listening disfluencies—based on the distance between the two. Previous research has shown that phonetic posteriorgram (PPG)-based DTW alignment between $L1_{S1}$ and $L1_{SS}$ can effectively derive word, syllable, and phoneme-level intelligibility annotations, as demonstrated in \cite{zhu2021multi, yue17_interspeech}.
\subsection{Seq2Seq Voice Conversion}
\label{seq2seqvc}
Conventional VC aims to change non-/para-linguistic features while preserving the linguistic content of input speech. However, with the advent of end-to-end architectures and self-supervised learning (SSL), recent studies on Seq2Seq VC enable a more robust mapping of sophisticated latent features between source and target voices. This includes tasks such as speaking style conversion \cite{maimon-adi-2023-speaking}, voice emotion conversion \cite{zhou_emotionVC}, and foreign accent conversion \cite{zhao_fac_journal, zhao_ppg}.

In \cite{huang20i_interspeech}, the authors first proposed a transformer-based VC model that enables sequential mapping between varying durations of source and target speech.
In \cite{hayashi2021non},  a non-autoregressive model was proposed using a conformer structure
address the issues of inaccurate duration prediction and repetitive artifacts caused by the auto-regressive nature of transformer models. Recent works have further improved upon this by incorporating Monotonic Alignment Search (MAS) and joint vocoder training, achieving superior performance in both duration and prosody \cite{okamoto2023e2e, huang2023aas}.

\subsection{Self Supervised Speech Representations}
Self-supervised speech representations (S3Rs) like Wav2vec 2.0\cite{wav2vec2}, HuBERT\cite{hsu2021hubert} and WavLM\cite{chen2022wavlm} achieved state-of-the-art performance on various speech benchmarks \cite{yangspeechfound}. Particularly, recent studies on unintelligible speech processing are signiciantly benefited from SSL representations,  such as dysarthric speech recognition \cite{wang2024unit} and electrolaryngeal speech enhancement \cite{lester_EL,violeta2024electrolaryngeal}. Additionally, self-supervised speech representation based VC (S3R-VC) has also improved the performance in both linguistic and acoustic aspects, even when only applied on the encoder side \cite{s3prl-vc-journal}.

In the following sections, these recent advances in Seq2Seq VC and S3R models are effectively adapted to implement our purposed L1-shadowing-L2 system.

\begin{figure}[t]
%     \vspace{-1cm}
	\centering
	\includegraphics[width=\columnwidth]{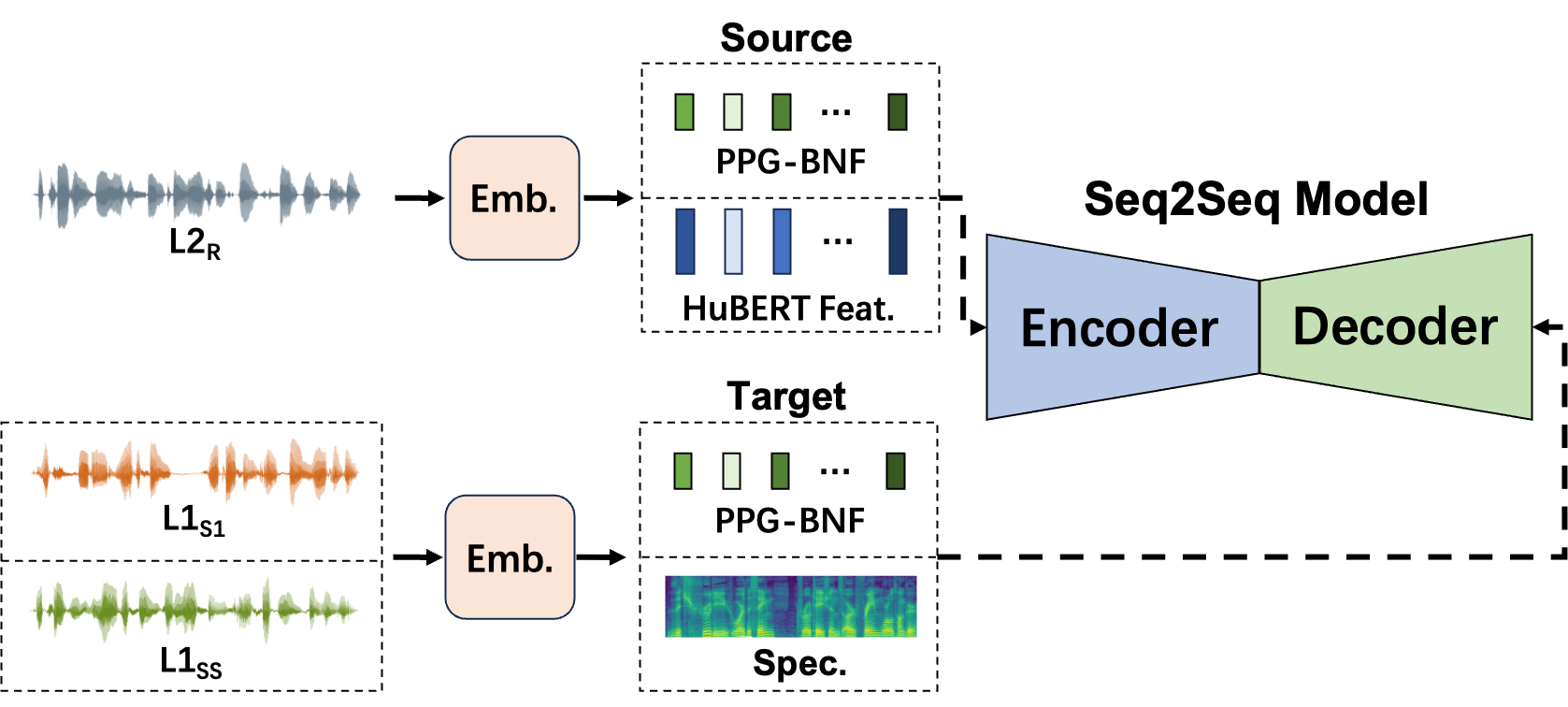}
    \vspace{-5mm}
 	\captionof{figure}{Overview of the proposed L1-shadowing-L2 system: after the source/target embeddings are processed by their respective encoders, the Seq2Seq model directly maps the source input $L2_{R}$ to the target outputs $L1_{S1}$ or $L1_{SS}$.
  }
  % Unlike previous methods using PPG-BNF and mel-spectrogram features, our approach leverages self-supervised feature, specifically, HuBERT feature, in the source phase to capture the general characteristics of L2 utterances.
 	\label{fig:vs_concept}
      \vspace{-5mm}
\end{figure}
\section{Purposed Method}

\subsection{Latent Speech Representation Embedding for L2 Speech}
Previous research in foreign accent conversion (FAC) has shown that bottleneck features extracted from phoneme recognition tasks (PPG-BNF) \cite{liu2021any} can be successfully applied in one-to-one voice conversion (VC) scenarios \cite{zhao_ppg, fac-evaluate}. In our previous work, we further demonstrated that this speaker-independent feature, PPG-BNF, can be effectively adapted to any-to-one scenarios \cite{geng2024APSIPA}.

Building on the success of S3Rs, we incorporated them into our study. As shown in Fig.~\ref{fig:vs_concept}, alongside PPG-BNF, HuBERT \cite{hsu2021hubert} features were selected as an alternative to represent L2 utterances. Although HuBERT is not explicitly designed to be speaker-independent, its strong representational capabilities are expected to capture features overlooked by PPG-BNF such as prosody, which can help identify unintelligible parts in L2 speech.

\subsection{Two-step Training Strategies for $L1_{S1}$-targeting System}
\begin{figure*}[t]
	\centering
	\includegraphics[width=\textwidth]{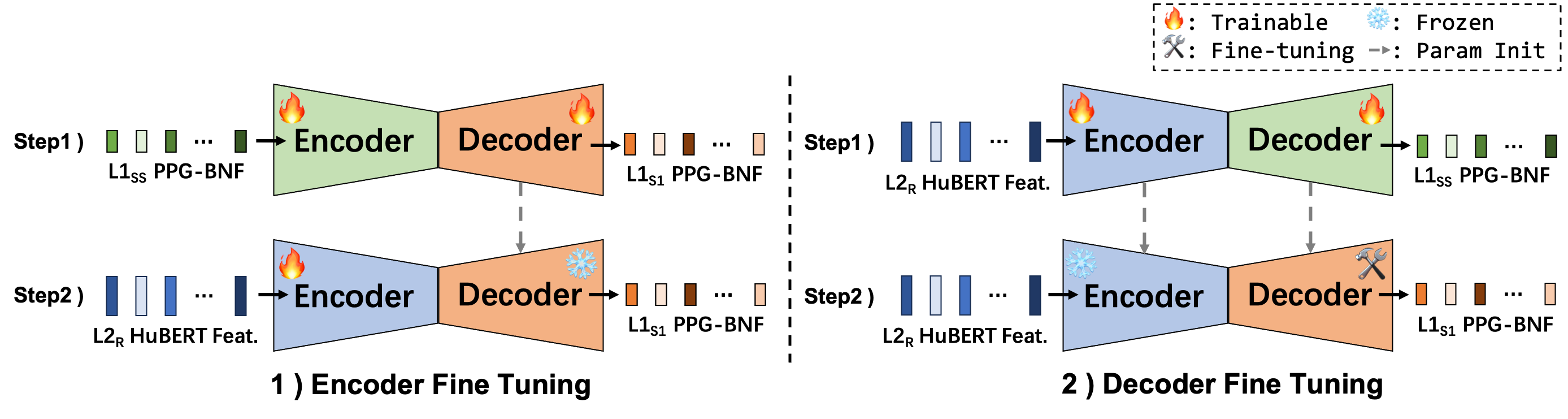}
	\captionof{figure}{Overview of two proposed fine-tuning approaches for mapping $L2_{R}$ to $L1_{S1}$ utilizing $L1_{SS}$.}
	\label{fig:FT_with_legend}
\vspace{-5mm}
\end{figure*}
\label{2_step}
As discussed in Sec.~\ref{seq2seqvc}, VC models are typically designed to preserve the linguistic content of input speech. However, in our study, the goal is not only to reduce the accent in L2 speech but also to highlight unintelligible segments. This means our model will intentionally modify the linguistic content. While directly mapping $L2_{R}$ to $L1_{S1}$ poses technical challenges, we propose two training strategies to fully exploit the similarities and differences between $L1_{S1}$ and $L1_{SS}$.

\subsubsection{Encoder Fine Tuning}
As illustrated on the left-hand side of Fig.~\ref{fig:FT_with_legend}, we draw inspiration from the autoencoder-style training method in \cite{vtn_journal, zhang2019non}, recognizing that the encoder in Seq2Seq models primarily extracts linguistic information. To capture the distance between shadowing and script-shadowing voices, we first use the PPG-BNF of $L1_{SS}$ and $L1_{S1}$ as the model’s source and target, respectively. Then, we fix the decoder parameters and reconstruct the encoder with the HuBERT features of $L2_{R}$.

\subsubsection{Decoder Fine Tuning}
As illustrated on the right-hand side of Fig.~\ref{fig:FT_with_legend}, we first train the model in a similar way to the FAC task. that is, $L2_{R}$ to $L1_{SS}$. Then we freeze the encoder and fine-tune the decoder using $L1_{S1}$ as the target. This approach aims to reconstruct the differences between $L1_{S1}$ and $L1_{SS}$ using the acoustic generation capabilities of the decoder.

\section{Experimental Settings}
\subsection{Dataset}
In this study, we used a reverse shadowing dataset based on \cite{yue17_interspeech, zhu2021multi}. Reading aloud utterances were collected from 225 Japanese English speakers ($L2_{R}$) with varying degrees of accent. A native male English speaker then shadowed and script-shadowed these recordings, producing $L1_{S1}$ and $L1_{SS}$, respectively, where 2,695 triplets of \{$L2_{R}$, $L1_{S1}$, $L1_{SS}$\} were prepared. Each dataset contained an average of 3.9 hours of valid phonation, with a mean sentence duration of 5.0 seconds. 300 utterances were selected for the test set.
\subsection{Seq2Seq Model}
The inherent differences between the context $L2_{R}$ and $L1_{S1}$ can easily lead to attention failures, particularly with high-dimension features like S3Rs. To mitigate this, we use AAS-VC \cite{huang2023aas} for source-target mapping. AAS-VC applies monotonic constraints and uses the MAS algorithm for advanced duration prediction, further enhanced by a forward-sum loss based on connectionist temporal classification (CTC) alongside the primary L1 training loss. This approach ensures that AAS-VC generates a quasi-diagonal alignment between source and target sequences, even when they are partially divergent, which is highly suitable for our task.

\subsection{Pretrained Embedding and Decoding Models}
For feature embedding, in addition to the original PPG-BNF designed by \cite{9428161}, we introduced HuBERT-base model as an alternative latent feature and selected the output from the ninth layer for embedding. The framewise feature dimensions of PPG-BNF and HuBERT are 144 and 768, respectively. 
For the target phase, only PPG-BNF and mel-spectrograms were utilized. In our preliminary experiments, we conducted HuBERT as the target; however, the attention alignment failed to converge under those conditions.

For the PPG-to-Spec decoding, we trained a single-speaker PPG-to-Spec decoder following the implementation outlined in \cite{s3prl-vc-journal}. For the vocoder across all models, we employed HiFi-GAN \cite{kong2020hifi}. The input mel-spectrograms was set to 80 frequency bins with a stride of 20 ms, aligning with the temporal resolution of the HuBERT features. Both the PPG-to-Spec decoder and vocoder were trained exclusively using the L1 speaker’s utterances.

\section{Experimental Evaluation}
\label{exp_eva}

\begin{table*}
  \centering
\caption{Evaluation of the L1-shadowing-L2 system: \textit{ANA-SYN} refers to the analysis-synthesis results obtained using the PPG-BNF approach \cite{zhao_ppg}. \textit{Baseline-S1} and \textit{Baseline-SS} refer to baseline models for $L1_{S1}$ and $L1_{SS}$, respectively \cite{geng2024APSIPA}. Similarly, the proposed methods, \textit{PM-S1} and \textit{PM-SS}, perform direct transformations from $L2_{R}$ to $L1_{S1}$ and $L1_{SS}$, using HuBERT embedding.}
  \begin{tabular}{@{}cccccccccc@{}}
  \toprule
  Description& Input     &Source    &Source feat.& Target    &Target feat.& S1-CER $\downarrow$ & S1-WER$\downarrow$ &UTMOS$\uparrow$&SpeechBERTScore$\uparrow$ \\
  \midrule
& $L1_{S1}$   &-   &-   & -  &-   & - & -   &4.04&- \\
& $L1_{SS}$   &-   &-   & -  &-   & 6.74& 12.46&4.21&0.840 \\
& $L2_{R}$        &-   &-   & -  &-   & 12.70& 24.25&3.09&0.618 \\
 \midrule
\textit{ ANA-SYN}& $L1_{S1}$   & -   & PPG-BNF& -   & -   & 2.23& 5.82& 3.40&0.8913\\
 & $L1_{SS}$   & -   & PPG-BNF& -   & -   & 7.54& 13.72& 3.47&0.8081\\
 & $L2_{R}$        & -   & PPG-BNF& -   & -   & 22.94& 40.26& 2.95&0.6045\\
\midrule
\textit{Baseline-S1}& $L2_{R}$   &$L2_{R}$&PPG-BNF& $L1_{S1}$&Mel.& 21.66& 37.73&3.40&0.6716 \\
& $L2_{R}$   & $L2_{R}$&PPG-BNF&  $L1_{S1}$&PPG-BNF& 38.07&   58.69&3.58&0.6437 \\
  \midrule
\textit{Baseline-SS}& $L2_{R}$   &$L2_{R}$&PPG-BNF& $L1_{SS}$&Mel.& 22.54& 39.36&3.47&0.7212 \\
& $L2_{R}$   & $L2_{R}$&PPG-BNF&  $L1_{SS}$&PPG-BNF& 28.60&   46.77&3.80&0.6882 \\
\midrule
\textit{ PM-S1}& $L2_{R}$   & $L2_{R}$   & HuBERT& $L1_{S1}$& Mel.& 18.02& 31.82& 3.70&0.7424 \\
 & $L2_{R}$   & $L2_{R}$   & HuBERT& $L1_{S1}$& PPG-BNF& 19.19& 33.53& 3.93&0.7568 \\
 \midrule
\textit{PM-SS}& $L2_{R}$   & $L2_{R}$   & HuBERT& $L1_{SS}$& Mel.& \textbf{16.31}& \textbf{28.53}& 3.74&0.7536 \\
 & $L2_{R}$   & $L2_{R}$   & HuBERT& $L1_{SS}$& PPG-BNF& 17.79& 31.67& \textbf{3.99}&\textbf{0.7651} \\
% \midrule
%  & $L2_{R}$   & $L1_{SS}$& PPG-BNF& $L1_{S1}$& PPG-BNF& 27.45& 47.68& 3.43&0.6716 \\
% & $L2_{R}$   & $L1_{SS}$& PPG-BNF& $L1_{S1}$& Mel.& 21.66& 37.73& 3.43&0.6716 \\
%  Method C)& $L2_{R}$   & $L2_{R}$   & HuBERT& $L1_{S1}$& Mel.& 25.80& 43.19& 3.44&0.6927 \\
%  Method C)& $L2_{R}$   & $L2_{R}$   & HuBERT& $L1_{S1}$& PPG-BNF& 20.70& 35.83& 3.91&0.7429 \\
%  Method D)& $L2_{R}$   & $L2_{R}$   & HuBERT& $L1_{S1}$& Mel.& 16.59& 29.74& 3.77&0.7579 \\
% Method D)& $L2_{R}$   & $L2_{R}$   & HuBERT& $L1_{S1}$& PPG-BNF& 18.33& 32.12& 3.96&0.7627 \\
\bottomrule
\end{tabular}

\label{tab:vs_latent}
\end{table*}
\begin{table}
  \centering
\caption{Evaluation for the 2-step strategies for converting $L2_{R}$ to $L1_{S1}$. }
  \begin{tabular}{@{}ccccc@{}}
  \toprule
  Description&Target feat.& S1-WER$\downarrow$ &UTMOS$\uparrow$&SpeechBERTScore$\uparrow$ \\
\midrule
\textit{ Enc. FT}& Mel.& 43.19& 3.44&0.6927 \\
        & PPG-BNF& 35.83& 3.91&0.7429 \\
 \midrule
\textit{ Dec. FT}& Mel.& \textbf{29.74}& 3.77&0.7579 \\
        & PPG-BNF& 32.12& \textbf{3.96}&\textbf{0.7627} \\
\bottomrule
\end{tabular}

\label{tab:vs_2step}
\end{table}

\subsection{Evaluation Metrics}
We focused on two aspects to evaluate our virtual shadowing system: linguistic similarity to practical $L1_{S1}$ and naturalness of generated speech. In particular, we introduced self-supervised feature based methods for convincing assessment\footnote{We carefully selected SSL models to ensure unbiased assessment. While HuBERT was used in our purposed methods, the ASR system for WER evaluation was based on wav2vec 2.0, and SpeechBERTScore was evaluated using WavLM.}.
\subsubsection{Linguistic similarity}
\label{speech_ling_sim}
Since our proposed virtual shadower aims to construct potentially corrupted speech generated by the L1 shadower, the output is different from what the L2 learner actually intends, but expects to be more similar to what the shadower actually uttered. Thus, 
in this study,
% while the WER referring $L2_{R}$'s ground truth is calculated as follows:
% \begin{equation}
% \text{WER}= \frac{S + I + D}{N_{R}},
% \end{equation}
we introduce S1-WER to evaluate the word-level linguistic similarity between the converted results and $L1_{S1}$:
\begin{equation}
\text{S1-WER}= \frac{\hat{S} + \hat{I} + \hat{D}}{N_{S1}},
\end{equation}
where $\hat{S}$, $\hat{I}$, and $\hat{D}$ denote for the counts of substitution, insertion, and deletion errors of converted speech referring to $L1_{S1}$'s ASR result, respectively. $N_{S1}$ is the word count of $L1_{S1}$ recognition transcript generated by the same ASR system. S1-CER is the character-based error rate, calculated in a similar  way to S1-WER.

\subsubsection{Naturalness}
\label{speech_nat}
We introduced UTMOS\cite{saeki2022utmos}, an automatic speech naturalness assessment method that can evaluate synthesized speech in mean opinion score (MOS) style. Furthermore, since the reference speech $L1_{S1}$ is available in our study, we also adopt a novel metric, SpeechBERTScore \cite{saeki2024spbertscore}, to evaluate the quality of the L1-shadowing-L2 speech. Let $\hat{Z} = (\hat{z}_n \in \mathbb{R}^D | n = 1, \cdots, N_{\text{gen}})$ and $Z = (z_n \in \mathbb{R}^D | n = 1, \cdots, N_{\text{ref}})$ denote the SSL speech features of the generated L1-shadowing-L2 speech and $L1_{S1}$. SpeechBERTScore is defined as a precision score:
\begin{equation}
\text{SpeechBERTScore} = \frac{1}{N_{\text{gen}}} \sum_{i=1}^{N_{\text{gen}}} \max_j \cos(\hat{z}_i, z_j),
\end{equation}
where $\cos(\cdot)$ is the pairwise cosine similarity between two features. Since S3Rs can capture rich linguistic and acoustic information, we expect SpeechBERTScore to evaluate both semantic similarity and naturalness of the generated speech.

% \input{nat}

% In addition to word-/character-level evaluation, BERTScore \cite{zhang2019bertscore}, which calculates the pairwise cosine similarity and inverse document frequency (idf) score between two token sequences, assesses their semantic similarity.

% BERTScore is more suitable for our task, An ideal shadowing sentence does not need to perfectly replicate the exact content but should preserve the intended meaning. While shadowing disfluencies like synonyms words may significantly affect word-level similarity, they have less impact on the overall sentence meaning. 
% Textual embedding like BERT with similar token representation for related words can levelage this distance.

\subsection{Analysis on Latent Representation Embedding}
As shown in Table~\ref{tab:vs_latent}, the purposed methods that incorporated S3R significantly improved both the linguistic similarity and the quality of the converted speech compared to the \textit{Baseline-S1/SS}. Specifically, the best results were achieved with \textit{PM-SS}, where S1-WER decreased from 37.73\% to 28.53\%, and SpeechBERTScore increased from 0.7212 to 0.7651, representing improvements of 24.4\% in linguistic similarity and 6.1\% in speech quality. In this FAC-like scenario, the L1 shadower fluently repeated the content intended by the L2 learner while viewing the learner’s script, creating favorable conditions for training. The reduction in S1-WER/CER further demonstrated the capability of S3Rs to reduce mispronunciation. 

For \textit{PM-S1}, the results surpassed those of \textit{Baseline-S1}, with S1-WER decreasing to 31. 82\% and improvements were observed in both UTMOS (3.93) and SpeechBERTScore (0.7568). However, the slightly distorted outcome compared to \textit{PM-SS} suggests that the model struggled with this challenging task, where both the linguistic content and the acoustic characteristics needed to be transformed. 

For target feature selection, both \textit{Baseline} and \textit{PM} models consistently exhibited better linguistic fidelity when mel-spectrograms were used as the target feature. Meanwhile, improvements on UTMOS, reflecting speech naturalness, highlight the distinct advantage of using PPG-BNF as the target feature. This is further supported by SpeechBERTScore, where models targeting PPG-BNF most closely resemble real $L1_{S1}$ in the latent space.

\subsection{Discussion on Two-step Training Strategies}
\label{discussion}
% As discussed in Sec.~\ref{2_step},  L1-shadowing-L2's output should resemble $L1_{S1}$ rather than $L1_{SS}$.
% Table~\ref{tab:vs_2step} presents our results using two-step training strategies targeting $L1_{S1}$. where the decoder fine-tuning method achieved the best performance among the four different settings. And also surpass the PM-A in Table~\ref{tab:vs_latent} which is the direct mapping from $L2_{R}$ to $L1_{S1}$. Encoder fine-tuning, however, showed no improvement compared to the baselines, indicating that the auto-encoder style training in this study requires further exploration. Although none of the proposed two-step training approaches surpassed the best result targeting $L1_{SS}$ in Table~\ref{tab:vs_latent}, 
As discussed in Sec.~\ref{2_step}, the L1-shadowing-L2 output should resemble $L1_{S1}$ rather than $L1_{SS}$. In this section, we explore the impact of our proposed two-step approaches which initialize the Seq2Seq model with $L1_{SS}$ and then target $L1_{S1}$.

Table~\ref{tab:vs_2step} presents the results of applying two-step training strategies aimed at $L1_{S1}$. Between the two approaches, \textit{Dec. FT} achieved superior performance, surpassing \textit{PM-S1} in Table~\ref{tab:vs_latent} and closely resembling the best result, \textit{PM-SS} across all metrics. \textit{Enc. FT}, however, showed no significant improvement over \textit{Baseline} methods, suggesting that the autoencoder-style training used in this study requires further investigation.

Since none of the proposed two-step training approaches outperformed \textit{PM-SS} in Table~\ref{tab:vs_latent}, we believe that relying solely on a deep learning model to implicitly capture the difference between $L1_{SS}$ and $L1_{S1}$ is insufficient. Furthermore, the results suggest that if the inherent pronunciation differences between $L1_{SS}$ and $L2_{R}$ can effectively reflect the ability of an L1 listener to comprehend L2 speech, a shadowing system could be developed without the need for real shadowing speech. Instead, parallel datasets, which are more feasible and easier to collect, could serve as a viable alternative.

\section{Conclusion and future work}
% In this study, we developed a virtual L1-shadowing-L2 system utilizing VC techniques to reduce accentedness and identify unintelligible segments of L2 speech. By employing latent speech representations like SSL and semi-parallel datasets, the system demonstrates promising performance in reconstructing L1-shadowed speech, maintaining both linguistic fidelity and speech quality. This novel application of VC methods offers significant potential, and we anticipate that the system will make valuable contributions to the field of language learning.

In this study, we developed a virtual L1-shadowing-L2 system to help L2 speakers identify unintelligible segments in their speech by providing feedback that replicates L1 shadowing behavior. By leveraging latent speech representations, particularly self-supervised speech representations (S3Rs), the system demonstrates strong performance in reconstructing L1-shadowed speech, preserving both linguistic fidelity and speech quality. However, the optimal method for utilizing L1-shadowing-L2 data remains an open question. 
Future work will focus on incorporating the distance between $L1_{SS}$ and $L1_{S1}$ as prior information. Specifically, we plan to integrate metrics such as PPG distance \cite{zhu2021multi} into the loss function, with appropriate thresholds to fully utilize listening breakdown data.

Additionally, experimental evaluation in Sec.\ref{exp_eva} demonstrated that using parallel data can also help to establish a promising shadowing system for specific speaker-listener pairs. 
While this study focused on L2 and L1 speaker interactions, global English communication often involves L2 speakers from different linguistic backgrounds.  As discussed in \cite{tomita24_interspeech}, creating a mutual shadowing system among L2 speakers using parallel L2 data could help learners better understand how comprehensible their L2 speech is.
Given the availability of parallel L2 corpora, such as \cite{zhao2018l2arctic}, we believe the implementation of such a system is feasible and expect it to contribute to global communication within the context of World Englishes.

\color{black}
\bibliographystyle{IEEEtran}
\bibliography{mybib}

%\end{thebibliography}

\vspace{12pt}

\end{document}